\documentclass[usenatbib]{mn2e}

\usepackage{epsfig,colordvi}
\usepackage{natbib}
\usepackage{color}

\usepackage{amssymb}
\usepackage{amsmath}

\newcommand{\be}{\begin{equation}}
\newcommand{\ee}{\end{equation}}
\newcommand{\bea}{\begin{eqnarray}}
\newcommand{\eea}{\end{eqnarray}}

\newcommand{\sci}[2]{#1$\times$10$^{\text{#2}}$}

\def \der{{\rm d}}

\def \r1r2{{|{\bf r}_{1}-{\bf r}_{2}|}}

\begin{document}

\title[Dark matter cascade decay]{Implications of dark matter cascade decay from DAMPE, HESS, Fermi-LAT and AMS02 data}

\author[Gao \& Ma]{Yu Gao$^{1,\star}$, Yin-Zhe Ma$^{2,3,\dagger}$ \\
$^{1}$Key Laboratory of Particle Astrophysics, Institute of High Energy Physics,
Chinese Academy of Sciences, Beijing 100049, China \\
$^2$School of Chemistry and Physics, University of KwaZulu-Natal, Westville Campus, Private Bag X54001, Durban, 4000, South Africa.\\
$^3$NAOC-–UKZN Computational Astrophysics Centre (NUCAC), University of KwaZulu-Natal, Durban, 4000, South Africa.\\
emails: $^{\star}$gaoyu@ihep.ac.cn\,
$^{\dagger}$ma@ukzn.ac.za;\,}

\maketitle

\begin{abstract}
Recent high-energy cosmic $e^\pm$ measurement from the {\it DArk Matter Particle Explorer} ({\it DAMPE}) satellite confirms the deviation of total cosmic ray electron spectrum above 700-900 GeV from a simple power law. In this paper we demonstrate that the cascade decay of dark matter (DM) can account for {\it DAMPE}'s TeV $e^+e^-$ spectrum. We select the least constraint DM decay channel into four muons as the benchmark scenario, and perform an analysis with propagation variance in both DM signal and the Milky Way's electron background. The best-fit of the model is obtained for joint {\it DAMPE}, Fermi-Large Area Telescope (Fermi-LAT), High Energy Stereoscopic System (HESS), high energy electron data sets, and with an $\mathcal{O}(10^{26})$ second decay lifetime, which is consistent with existing gamma ray and cosmic microwave background limits. We compare the spectral difference between the cascade decay of typical final-state channels. The least constrained $4\mu$ channels give good fits to the electron spectrum's TeV scale down-turn, yet their low energy spectrum has tension with sub-TeV positron data from AMS02. We also consider a three-step cascade decay into eight muons, and also a gamma-ray constrained $4\mu,4b$ mixed channel, to demonstrate that a further softened cascade decay signal would be required for the agreement with all the data sets. 
\end{abstract}


\section{Introduction}
\label{sec:intro}
The {\it DArk Matter Particle Explorer} ({\it DAMPE}) satellite releases recent data of the combined electron and position energy spectrum from ${\cal O}(10)$ GeV to 10 TeV in energy~\citep{DAMPE}. The observed spectrum confirms H.E.S.S.~\citep{HESS2017} observation of a change of spectral shape at around 700-900 GeV. Below this energy the spectrum is close to a power law which is in agreement with {\it Fermi-LAT}~\citep{bib:fermi_e} and H.E.S.S. The spectrum drops quickly in the higher energy regime. 

The cosmic $e^\pm$ spectral changes may rise from astrophysics or new physics sources at the TeV scale. The Universe's Dark matter (DM) can be consist of hypothetical, unseen particles. Such DM particles can couple to normal matter, and annihilate or decay into the Standard Model's (SM) particles~\citep{Feng10}. The Galactic DM halo has been widely studied in attempt to account for the cosmic position fraction observation from {\it PAMELA} telescope~\citep{bib:pamela} and later by {\it Fermi-LAT}~\citep{bib:fermi_e}, and {\it Alpha Magnetic Spectrometer} (AMS02)~\citep{bib:ams02} to high precision, as well as an increase in the cosmic anti-protons~\citep{bib:pamela_antiproton,bib:ams02_antiproton}. Alternatively, astrophysical sources such as pulsars~\citep{bib:pulsar,bib:Yuksel09,bib:Shaviv09,bib:Biermann09,bib:Blasi09,bib:Malyshev09} and supernova remnants~\citep{bib:remnants} are often studied as the potential sources for the positron excess. Interestingly, changes in spectral shape, indicating for a more sophisticated origin, are also observed in proton and light nuclei (see ATIC~\citep{bib:atic} and AMS02~\citep{bib:ams02_antiproton}).

The dark matter hypothesis faces many constraints. DM annihilation and decay inject high-energy $e^\pm$ and photons. The prompt photons, along with inverse Compton radiation off injected $e^\pm$, are constrained by diffuse and point-source gamma ray searches.  In case of annihilation, the annihilation rate can be enhanced by concentrated DM distribution at the Galactic center or in sub-halos. Gamma ray measurements for the Galactic Center (GC) and dwarf spheroidal galaxies (dSph) from {\it Fermi-LAT}~\citep{bib:fermi_gamma} and H.E.S.S.~\citep{HESS2017} provided significant bounds on DM annihilation signals. 

Another important bound of annihilation comes from the reionization history. DM-originated $e^+e^-$ and photons can ionize neutral hydrogen atoms after the recombination, and consequently alter the propagation of the cosmic microwave background (CMB). {\it Planck} 2015 data offers a severe limit on DM's zero-velocity annihilation rates~\citep{bib:planck}. 

In the DM decay scenario, the strength of signal depends on the total DM abundance instead of its central density profile, thus the decay's gamma ray signal is less manifest from the concentration in galactic halos, hence less constrained by gamma rays. Recent studies (e.g.~\citet{Liu:2016ngs}) based on DM gamma-ray emissions in the extragalactic diffuse photons showed that DM decay channels can explain the position excess within {\it Fermi-LAT}'s limits. CMB constraint from {\it Planck} is also much weaker for DM decay due to a different redshift dependence in comparison to annihilation. Recent studies (\citet{Slatyer:2016qyl, 2018PhRvD..98d3006C}) of DM impacts on the CMB temperature and polarization anisotropy showed that a DM decay lifetime greater than ${\cal O}(10^{24-25})$ second is allowed by {\it Planck} measurement.

Decaying TeV-scale dark matter readily arise as metastable particles in theories Beyond the Standard Model (BSM) with a lifetime significantly longer than the age of Universe. Typical scenarios include the hidden-sector's particles, which in general can be destabilized due to UV-scale physics mediation~(\cite{2009PhRvD..79j5022A}) and become suitable decaying dark matter candidates. Extended gauge symmetries motivate for heavy right-handed neutrinos, which may also acquire a long lifetime via flavor structures~(\cite{2009PhRvD..80g3017A}), etc. Such metastable DM decay can provide a viable source for the TeV-scale cosmic ray electron and positron data.

In this paper, we adopt a DM cascade decay scenario, in which multiple decay steps soften the source $e^+e^-$ spectrum, making it suitable to fit the {\it DAMPE}'s electron data while being consistent with {\it Fermi-LAT} and H.E.S.S. measurements. We consider a variant power-law Galactic electron background in the presence of DM decay signals. In the following, we discuss the optimal DM decay channels and cascades with more than two decay steps in Sections~\ref{sect:channel} and~\ref{sect:3step}. Electron propagation and the Galactic background are discussed in ~\ref{sect:prop}. Then we demonstrate the fitting result to joint {\it DAMPE, Fermi-LAT, HESS and AMS02} data with our cascade decay model in Section~\ref{sect:fit}, and conclude in Section~\ref{sec:conclude}.

\section{Dark matter cascade decays}
\label{sect:channel}

Dark matter in the Galactic halo decays at a constant rate that is insensitive to the dark matter particle velocity and small-scale density distribution of the density profile. With a small decay width $\Gamma$, DM decay injects SM particles at a steady rate, 
\be 
\frac{\der^{2} N}{\der E \der V} = \Gamma \frac{\rho}{m_{\rm DM}} \frac{\der N}{\der E},
\ee
where $\rho$ is the dark matter density, and the lifetime $\tau_{\text{DM}}=\Gamma^{-1}$ should be much longer than the age of Universe. ${\der N}/{\der E}$ is the SM particle's injection spectrum. To produce TeV electrons, the dark matter particle mass needs to be multiple TeV or heavier. The photons and neutrinos can propagate to the Earth uninterruptedly. The charged particles ($e^\pm, p, \bar{p}$) are affected by the Galactic magnetic field and diffuse away from the source in a random walk motion. 

The shape of $e^\pm$ injection spectrum ${\der N}/{\der E}$ is model-dependent and can differ dramatically among different decay channels. So far, the most precise sub-TeV $e^\pm$ measurement from AMS02 empirically require the post-propagation spectrum to have power-law index near -2.7. This spectral shape provides guidance that the DM's $e^+$ injection spectrum needs to be soft. In addition, to avoid an abrupt end-point in the combined $e^\pm$ spectrum, the injection spectrum to be softened near the high-energy end.

If DM directly decays into $e^{+}e^{-}$ pairs, the injected spectrum is monochromatic. The monochromatic $e^\pm$ may help to explain {\it DAMPE}'s `feature' at 1.4 TeV~\citep{Fang:2017tvj,Yuan17,Fan17,Duan17,Chao:2017emq}, yet its shape is too hard to fit the positron data. A common solution is that the decay goes through intermediate decay steps before electrons emerge. Such a process is referred as `cascade' decay. Depending on the number of immediate steps, the final electrons spectrum can vary from being as hard as near-monochromatic, to very soft as that in hadronic showers.

\begin{figure}
\includegraphics[scale=0.6]{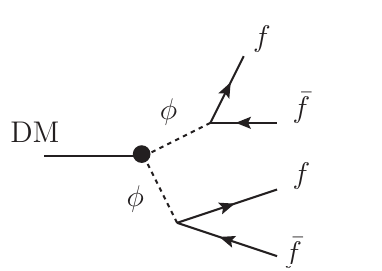}
\caption{The DM decays into a pair of mediators that subsequently decay into a four-lepton final state. 
}
\label{fig:feynman}
\end{figure}

In this analysis, we adopt the cascade decay DM$\rightarrow 2\phi\rightarrow 4\mu$ as the benchmark scenario to test the {\it DAMPE} data. As illustrated in Fig.~\ref{fig:feynman}, the DM first decays into a pair of mediators that subsequently decay into four SM fermions. The fermion's spectrum depends on the relative size of DM and $\phi$ masses, which classifies into two typical cases: (1) a `heavy mediator' (HM) case that $M_\phi\sim M_{\text{DM}}$, where the $\phi$ pair is produced non-relativistically and each fermion carries energy equal to $M_\phi/2\approx M_{\text{DM}}/4$. This HM scenario can be simplistically related to a two body decay process DM$\rightarrow f\bar{f}$ with half the DM mass. While the final state fermion kinematics are the same, the DM mass and fermion multiplicity are both doubled, leading to an identical signal rate and best-fit decay life-time. (2) a `light-mediator' (LM) case with $M_\phi \ll M_{\text{DM}}$, where $\phi$ is highly relativistic and its large Lorentz boost makes the lab-frame spectrum of $f$ into a plateau shape. This softens the $f$ spectrum, and its decay products are further softened. The HM spectrum has a hard injection spectrum and suffers from stronger gamma ray constraints, and we only consider the LM scenario in this work.

The LM scenario has two main benefits. First, with a low $\phi$ mass the decay channel of $\phi$ can be controlled by its relative size to SM fermion masses. A scalar $\phi$ can preferably decay into the heaviest fermion if kinematically allowed, giving preference for a specific channel. Alternatively, flavor coupling structure can also be introduced to enhance the branching ratio into particular fermion(s), for instance, via heavy non-SM gauge fields that generate a $\phi\rightarrow f\bar{f}$ at loop level~\citep{Allahverdi:2009ae}. As the the signal rate is insensitive to the $\phi$ lifetime, for Galactic DM sources, a $\phi$ of a few GeV mass can easily evade collider and direct detection bounds if its to SM fermions are small. Second and more importantly, if $M_\phi$ is closer to the fermion-pair mass threshold, a limited $Q^2$ in $\phi$ decay suppresses photon radiation, and reduces the constraint from diffuse gamma rays, e.g. in secluded DM scenarios~\citep{Profumo:2017obk}, etc. With these considerations, we adopt the LM four-muon channel, where a $M_\phi$ close to $2M_\mu$ can provide a large 4-muon decay branching ration and reduce photon radiation. In comparison, a direct DM$\rightarrow 4e$ cascade still has a hard spectral end-point. The DM$\rightarrow 4\tau$ channel suffers from $\pi^0$-decay photons that cannot easily evade extragalactic gamma ray bounds ~\citep{Liu:2016ngs}. Since the DM decay signal only depends on DM lifetime and the total DM quantity, the extragalactic bound for decaying DM is insensitive to assumptions in halo  sub-structures.

\begin{figure}
\includegraphics[scale=0.5]{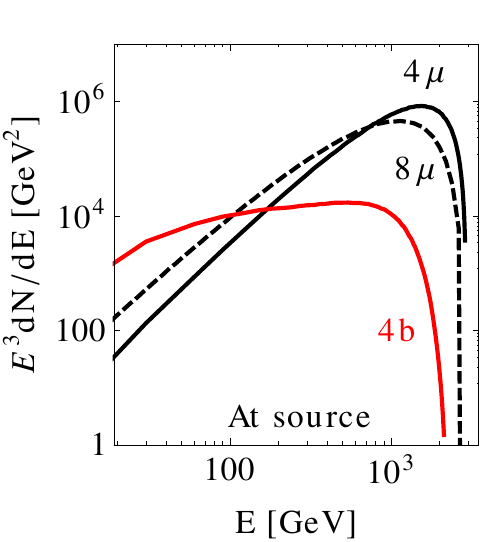}
\includegraphics[scale=0.5]{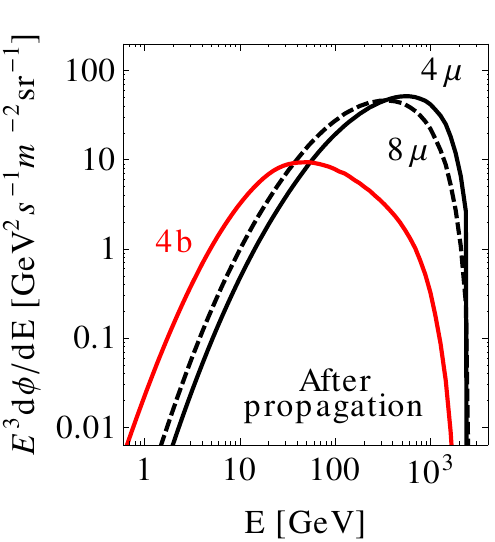}
\caption{{\it Left}-- Prompt injection $e^\pm$ spectra prior to propagation, from four-body DM cascade decays into muons, $b$-quarks and a three-step cascade into eight muons. {\it Right}-- Differential flux intensity after propagation to the Earth, the spectra are softened due to energy loss. For the four-body cascades, $M_{\text{DM}}=6$\,TeV and $M_\phi/M_{\text{DM}}\ll 1$. The three-step cascade assumes twice the DM mass than four-body channels. Due to propagation modeling uncertainties, here each propagated spectrum assumes its own channel's best-fitting {\sc galprop} parametrization to the $e^\pm$ data with an Einasto dark matter halo profile.}
\label{fig:injection_spec}
\end{figure}

Fig.~\ref{fig:injection_spec} illustrates the prompt spectra right after DM decay (at source) and the differential flux after propagating to the Earth. For comparison $4b$ cascade channels is shown as a representative of typically much softened spectra from hadronic cascade channels. Noted that the $e^\pm$ propagation effects strongly depends on the modeling of the cosmic ray propagation, where the particle diffusion speed and energy loss depend on the assumptions of the Galactic radiation and magnetic fields. In practice, the propagation and background parametrization are determined by fitting the Galactic and DM injection spectra to the $e^\pm$ data (see Section~\ref{sect:prop} and ~\ref{sect:fit} for details). Note that while the $e^\pm$ data can pin down the parameter values in the fit with a particular channel's injection spectrum, the signal's spectral shape varies significantly between different DM decay channels, leading to variation of best-fit propagation models between signal channels. For instance, if we adopt the best-fit propagation parameters from the ($4b-4\mu$ mixed) channel, the lowest energy part of $4\mu$ spectrum around 100 GeV may rise by a factor $\sim$1-2 in comparison to that using the $4\mu$ channel's own best-fit propagation parameters.

Fig.~\ref{fig:injection_spec} shows the `best-fit' propagated $e^\pm$ spectra by fitting each channel's injection spectrum and the propagation parameters to the {\it DAMPE}, FermiLAT, AMS02 and H.E.S.S. data. 
A much softened $e^\pm$ spectrum from a hadronic channels is also show-cased for comparison. The hadronization process in hadronic channels and following soft cascade hadron decays leads to multiple decay steps, which effectively softens the electron injection spectrum. However, hadronization also produces numerous neutral pions ($\pi^{0}$) that would cause tension in diffuse gamma-ray observation.

\section{Further cascades}
\label{sect:3step}

Generally the final-state electron energy softens with an increased number of intermediate decay steps in the cascade. While the previously discussed two-step cascade decay readily demonstrates the softening on the electron spectrum, we may pursue it further by considering more decay steps in this process. For TeV dark matter theories, a three-step cascade can rise in the case when the heavy mediator, for instance a scalar that breaks a non-SM gauge symmetry, primarily decays into the non-SM gauge bosons if kinematically permitted. The gauge bosons would then decay into the SM fermions, leading to two intermediate decay steps.

As successful softening requires the `LM' scenario, in each decay step the daughter particle's mass should be significantly lower than that of its parent. This condition kinematically prevents inserting a large number of intermediate steps: for a TeV-scale dark matter to produce a softened injection spectrum in the GeV range, three to four decay steps would saturate the GeV-TeV gap if each parent particle mass is at least one order of magnitude higher than its daughter mass. 

It is also interesting to see whether a chiral coupling between the mediators and the SM muons can help softening the electron injection spectrum. Generally a relativistic right-handed muon decays to lower energy electron in comparison to a left-handed one. Although a right-handed chiral coupling guarantees right-handed muons in the center of mass frame, the parent particle would still have carried a Lorentz boost up to $M_{\rm DM} /M_n$ that can flip the `backward' going muon's helicity after boosting back to the lab frame. Here $M_n$ denotes the mediator mass in the $n$th decay step. In comparison the `forward' muons along the boost will maintain their helicity. This leads to a higher fraction of `forward' muons, which fill up the higher-energy part of the muon spectrum, to keep the same chirality with the mediator-muon coupling. The lower-energy spectrum would demonstrate a larger fraction of the opposite chirality as it mostly consists of the `backward' muons with their helicity flipped by the boost. The combined effects with a right-handed chiral coupling lead to softening that is more visible near the high-energy end in the electron spectrum after muon decay. The low-energy end of the electron spectrum, however, does not show significant improvement over that from an unpolarized muon injection case.

The electron spectrum from a three-step cascade is shown ($8\mu$, dashed) in Fig.~\ref{fig:injection_spec} where the muons are statistically unpolarized. Compared to two-step decays, the entire spectrum shifts towards lower energy and the spectral shape in high energy part is significantly softened. At the low energy end, however, the spectral power-law indices are quite similar. This spectral similarity at very low energy end is mostly due to the finite mediator masses, which are an order of magnitude lower than the mass of their parent, but are still higher than the GeV scale. This reduces the phase-space for final-state muons to obtain energies much lower than the mediator mass. 

\section{$e^+e^-$ Propagation}
\label{sect:prop}

$e^\pm$ engage in diffusion motion inside the turbulent Galactic magnetic field. The propagation is described by the diffusion and energy-loss equation,
\be 
\frac{\der \Phi}{\der t}-D(E)\cdot \nabla^2\Phi-\partial_{E}(D_p(E)\cdot \Phi)=Q\,,
\ee
where $\Phi$ is the cosmic ray flux and $Q$ is the source term for astrophysical and DM injection. Electrons lose energy from synchrotron radiation, inverse Compton scattering and bremsstrahlung processes, and the spectrum soften over the propagated distance. We use the {\sc galprop} code~\citep{Strong:1999sv,bib:Strong01} for the numerical simulation of the propagation, in which the spatial diffusion coefficient $D(E)$ is parametrized~\citep{Strong:2007nh} as
\be
D =\beta D_0 \left({R \over R_0 }\right)^\delta \text{cm}^2\,\text{s}^{-1},
\label{eq:diffusion}
\ee
where $\beta\sim 1$ is the cosmic ray velocity. $R$ denotes the cosmic ray's rigidity. For $\delta$ we let it be greater than $1/3$ as it is required by a turbulent magnetic field with a Kolmogorov spectrum. The energy-loss coefficient $D_p$ is dynamically evaluated in the propagation process. Energy loss include bremsstrahlung, inverse Compton scattering on the CMB and the interstellar radiation field profile~\citep{Moskalenko:2005ng}, and synchrotron radiation loss in the cylindrical diffusion zone of 20 kpc in radius and 4 kpc in height, where the Galactic magnetic field strength is
\be 
B = B_{\odot} \exp\left( -\frac{r-r_\odot}{r_{\rm s}}\right) \exp\left(-\left|\frac{z}{z_{\rm s}}\right|\right)
\ee
with a local field strength $B_\odot=5\ \mu$G, and scales $r_{\rm s}=10$ kpc, $z_{\rm s}=2$ kpc.
For a reference rigidity $R_{0}\sim $~GV the diffusion parameter would take values in the range of $\left(3-5\right) \times 10^{28}\,{\rm cm}^{2}\,{\rm s}^{-1}$ to fit Galactic nuclei data~\citep{Strong:2007nh}. 

Assuming spatially constant values for the diffusion parameters ${D_0, R_0, \delta}$ in Eq.~(\ref{eq:diffusion}) does not fully account for spatial variation in Galactic magnetic fields. Anisotropic diffusion modeling has been shown to improve fitting the `Galactic center excess' (GCE)~\citep{Goodenough:2009gk,TheFermiLAT:2017vmf} in gamma rays for DM annihilation~\citep{Zhou:2014lva,Duerr:2015bea,Carlson:2016iis,Malyshev:2017pmr} where the annihilation rate is halo's density-square-enhanced near the Galactic center region. In our less enhanced DM decay case, Eq.~(\ref{eq:diffusion}) gives individually near-perfect fit to each experiment's data with degeneracy in these three parameters, and the main contribution to the joint $\chi^2$ arises from the difference between data sets as shown in the following section. Thus ignoring the spatial anisotropy in ${D_0, R_0, \delta}$ suffices for our fitting process.

For the DM signal we adopt the 
Einasto profile~\citep{Einasto65} for Galactic DM distribution to compute the $e^\pm$ injection source intensity,
\begin{equation}
\rho(r)=\rho_{\odot}\exp \left[-\frac{2}{\alpha}\left(\frac{r^{\alpha}-r^{\alpha}_{\odot}}{r^{\alpha}_{\rm s}} \right)  \right],
\label{eq:einasto}
\end{equation}
where we adopt the parametrization $\rho_{\odot}=0.3$ GeV\,cm$^{-3}$, $\alpha=0.17$, $r_{\rm s}=25$~kpc from~\citet{Cholis:2008vb}. For decaying dark matter, the choice of a cuspy or non-cuspy distribution leads to  comparable best-fit decay lifetimes, and the Einasto profile presents an adequate example for typical cold dark matter distribution. Noted that the non-cuspy isothermal profile~\citep{IsoT} can be more favored by recent N-body simulation results~\citep{bib:Bullock17} as it better explains the missing satellite problem~\citep{bib:Bullock17}. In case of decaying dark matter, the total decay rate depends on the halo mass and is insensitive to its distribution. The diffusion process would still depend on the source's spatial distribution, leading to relatively longer diffusion duration for cuspy profiles that are more distributed near the Galactic center. Compared to fitting with the Einasto profile, choosing the non-cuspy isothermal profile only leads to 1\% difference in the goodness of fit and a less than 10\% shift in the best-fit DM signal rate, thus it does not qualitatively affect the fit to TeV data.

Right panel in Fig.~\ref{fig:injection_spec} shows the DM's $e^{\pm}$ spectrum after propagation to the Earth. Comparing to the injection spectrum, the spectral shape becomes sufficiently softened and suitable to fit experimental data. After propagation the peak in each channel's spectrum locates at different energies, and the optimal DM mass can be inferred by aligning the peak to {\it DAMPE}'s turning point, i.e. $M_{\text{DM}}\sim 6$\,TeV for $4\mu$, 10 TeV for $4\tau$. The $4b$ cascade channel would require a $100$\,TeV-scale DM mass as its propagated peak energy locates at $O(10^{-2})$ of $M_{\text{DM}}$.

Noted that hadronization in DM$\rightarrow 4b$ and other hadronic channels would contribute to a signal in cosmic ray anti-protons. A recent study~\citep{Cui17} showed that in case of DM annihilation, the $b\bar{b}$ channel with a thermal cross-section improves the fit to antiproton measurements, and it is consistent with Fermi-LAT and {\it Planck} constraints. The DM decay case would have a less stringent diffuse gamma ray bound, thus its impact on the anti-protons can also be important to study. Here we do not pursue it further as the preferred $100$ GeV mass is insufficient to fit {\it DAMPE} results.

It is also interesting to discuss whether the cascade decay channels can account for the GCE in lower energy gamma rays. A recent study~\citep{Clark:2017fum} showed that the 4$\tau, 4b$ annihilations of DM in the sub-100 GeV mass range can help explain the GCE and AMS02's antiproton data. For a multi-TeV DM, the low energy tail of $4\mu$ channel's gamma ray injection spectrum lacks the spectral feature to fit the GCE. In addition, being a gamma ray poor channel, the photon emission in the $4\mu$ final state is model dependent and can be suppressed if the mediator mass approaches to the threshold of twice the muon mass. While 4$\tau$, 4$b$ cascade channels have softer gamma ray spectra, these two channels are readily excluded by extragalactic gamma ray bounds at the decay lifetime H.E.S.S. and {\it DAMPE} data require, plus the multi-TeV DM masses are also 1-2 orders of magnitude higher than the GCE preferred range~\citep{Clark:2017fum}. Therefore these channels may not make good candidates for the GCE at their best fit cases for the near-TeV `knee' in Galactic electron spectra.

\begin{figure}
\includegraphics[scale=0.55]{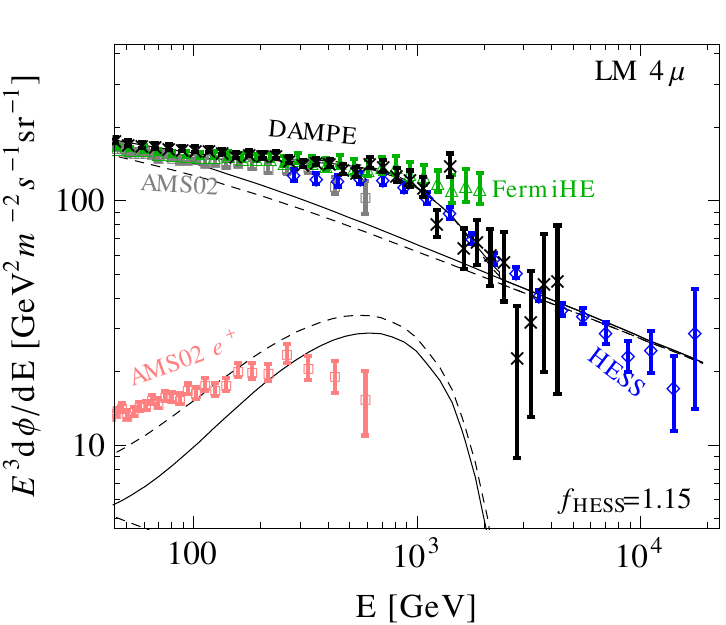}
\includegraphics[scale=0.55]{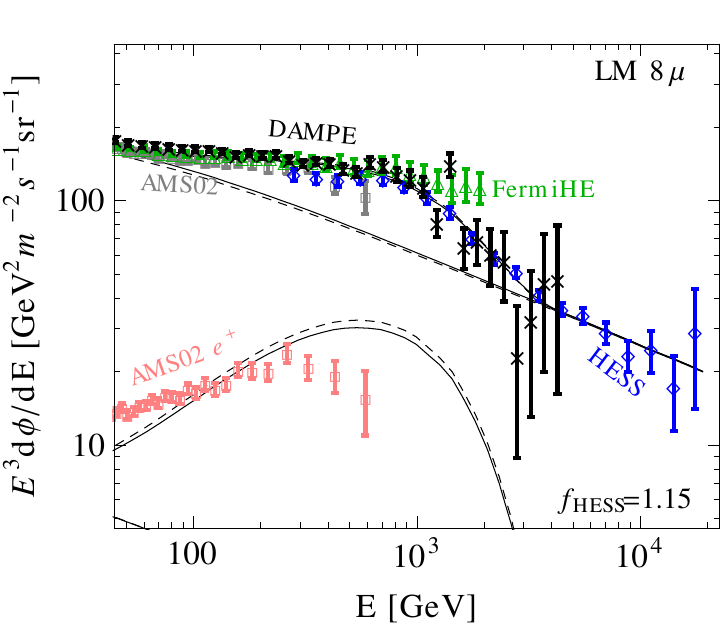}
\caption{Sample DM$\rightarrow 4\mu$ (upper) and three-step cascade into 8$\mu$ (lower) that give the lowest $\chi^2$ fits to the combined {\it DAMPE} and H.E.S.S. data, and the positron fraction data from AMS02 (see Table~\ref{tab:fit}). In each panel, the solid curves show the fit without including AMS02 data, while for the dashed curves AMS02 positron fraction data are included. Each set of three curves (top down) are the DM+Galactic $e^\pm$, Galactic $e^\pm$ and DM+Galactic $e^+$ spectra. $f_{\rm HESS}$ is the energy scaling factor for H.E.S.S. data.}
\label{fig:best_fit}
\end{figure}

\section{Fit to {\it DAMPE}, Fermi-LAT, H.E.S.S. \& AMS02}
\label{sect:fit}

We perform fitting to the joint {\it DAMPE}~\citep{DAMPE}, H.E.S.S.~\citep{HESS2017}, Fermi-LAT `high energy' (FermiHE)~\citep{Abdollahi:2017nat} and AMS02’s recent 30 million $e^{\pm}$ data~\citep{TingCernAMStalk}. For data selection, we focus on the high energy part of spectra and only include data above 80 GeV in each data set. Due to the power-law falling in the spectrum, lower energy data are measured with much higher statistics. The low energy spectrum can be sensitive to the modeling of astrophysical sources and solar modulation. Yet more importantly the high statistic weight in the low energy data can over-power the presence of a DM signal's spectral correction, if included in an evenly weighted analysis. With our main focus on the DM, we drop the data below 80 GeV to avoid sensitivities in low energy astrophysical modeling.

We use {\it DAMPE}, Fermi-LAT and and H.E.S.S. `total electron' data (both electron and positrons) in the fitting. For AMS02, note that its electron flux normalization differs from that of Fermi-LAT and {\it DAMPE}. All these  experiments have high-statistics data points below 1 TeV, and directly fitting the electron flux data would lead to large discrepancy between the data sets. As a workaround, we fit AMS02's positron fraction data instead of the absolute $e^{\pm}$ flux, which effectively removes the dependence on the normalization in AMS02 $e^\pm$ fluxes. 

H.E.S.S. data have large systematic uncertainty due to atmospheric hadronic modeling~\citep{HESS2017}. If naively added into quadrature, this systematic uncertainty overwhelms the statistical uncertainty and renders H.E.S.S. spectral shape ineffective in the analysis. Here we consider the systematic uncertainties to be correlated and adopt the energy scale factor $f$ that lets H.E.S.S. electron spectrum to shift in energy and magnitude, while keeping its shape unchanged. The H.E.S.S. contribution to the total $\chi^2$ would then only involve the scaled statistical uncertainty, plus an additional term that accounts for the scaling,
\be 
\chi^2_{\rm H.E.S.S.}=\sum\frac{(\phi_i^{\rm th}-f\cdot \phi_i^{\rm ex})^2}{f_{}^2\delta\phi_i^2} +\frac{(f-1)^2}{(\delta f)^2},
\label{eq:Hess_rescale}
\ee
where $\phi^{\rm th},\phi^{\rm ex}$ denote the theoretical and measured electron flux. $f$ is the energy scaling factor and we consider a typical $\delta f= 15\%$ that matches the range of H.E.S.S.'s systematic uncertainty\citep{bib:hess}. Due to differences between the data sets, the best-fit is found to prefer an upfloat around $f\sim 1.3$ for best agreement, which appears to be larger than the H.E.S.S. systematic uncertainty range. In the following results we will compare a restricted $f\le 1.15$ case, and also a case that $f$ is allowed to scale further.

The $\chi^2$ from each data set are summed together without additional weighting. {\it DAMPE}, Fermi-LAT high energy set, H.E.S.S and AMS02 have 29, 23,19 and 15 data points above 80 GeV. Note H.E.S.S. has one additional `data' point in Eq.~(\ref{eq:Hess_rescale}) after introducing the energy scaling factor $f$. 

\begin{figure}
\includegraphics[scale=0.55]{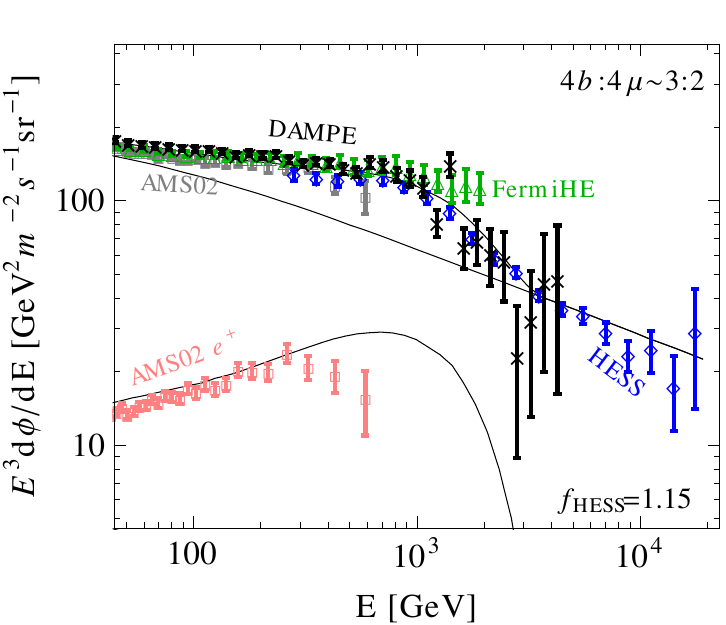}
\includegraphics[scale=0.55]{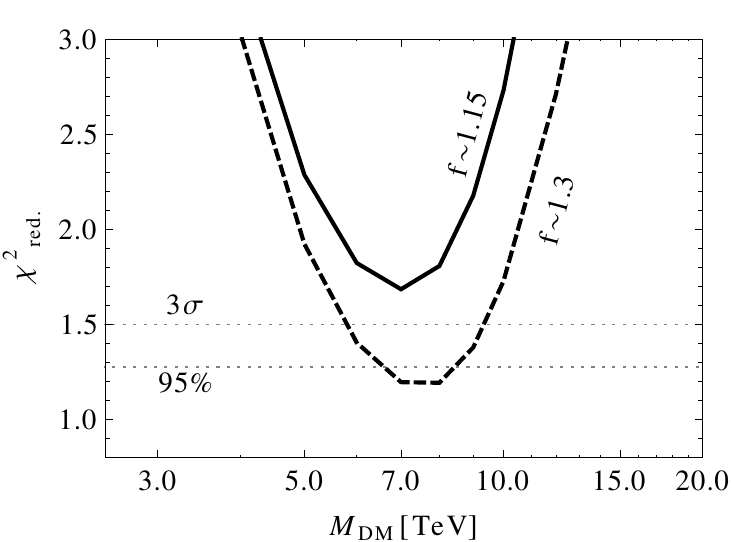}
\caption{Best-fit (upper) mixed $4\mu,4b$ channel for all four data sets. The goodness of fit (lower) is plotted with two ranges of H.E.S.S. scaling factor: $|f-1|\le 15\%$ (solid) and a free range (dashed) where $M_{\rm DM}=7-8$ TeV. $f\sim 1.3$ achieves the best agreement between datasets with an acceptable it ($\chi^2$=95 for 87 data points with 8 free parameters).
}
\label{fig:best_fit_mixed}
\end{figure}

We assume a single power-law spectrum for the Galactic electron background. The total cosmic ray spectrum consists of both the background and DM contributions,
\be 
\phi = \phi_{\rm bkg}(D_0, \rho_0, \delta, \phi_{0}, \delta_{\rm e}) + \phi_{\rm DM}(D_0, \rho_0, \delta),
\ee
where the background cosmic electron receives major contribution from supernova injection and minor contribution from isotope decays. $\phi_0$ is the normalization of Galactic electron flux at the reference rigidity 3.45GV, and $\delta_{\rm e}$ is the supernova injection power-law index above 4 GV. $\{D_0, \rho_0, \delta\}$ are the diffusion parameters in Eq.~(\ref{eq:diffusion}). For solar modulation, we adopt a fixed modulation potential at 700 MeV. 

Noted that more complicated Galactic spectral shape in cosmic rays are generally possible. For instance, a rising deviation from a simple power-law at several hundred GV in proton and light nuclei spectra, as measured by AMS02~\citep{Aguilar:2016kjl,Aguilar:2015ctt}, Advanced Thin Ionization Calorimeter (ATIC,~\cite{Panov:2011ak}), Cosmic Ray Energetics And Mass (CREAM, \citet{Yoon:2017qjx}) and other experiments. A Galactic broken power-law spectrum with a down-turn at TeV could reduce the requirement on DM contribution. However, such astrophysical hypotheses~\citep{Fang:2017tvj} will also need consolidation in future studies. In this work, we restrict to the simple power-law case for the DM hypothesis as it assumes the fewest degrees of freedom for the non-DM spectral components.

Thus our fitting parameter set is composed of the following:
\be 
\{ D_0, \rho_0, \delta; \phi_0, \delta_{\rm e}; f_{\rm HESS}, \tau(\Gamma_i) \},
\ee
where the first five parameters model the particle diffusion and the Galactic $e^\pm$ background. $f_{\rm HESS}$ represents the spectral shift in H.E.S.S. data due to energy scaling uncertainty, and $\tau$ is the DM decay lifetime that determines the magnitude of signal rate. Note $\tau$ may contain multiple (free) partial widths $\Gamma_i$ in case of mixed channels. Depending on the shape of the DM signal spectrum, the best-fit diffusion and background parameters may vary between different decay channels: for example, the Galactic electron spectral power index $\delta_{\rm e}=-2.56$ in a fit with 8 TeV $4\mu$ channel can shift to $\delta_{\rm e}=-2.62$ in a fit with the 8 TeV mixed $4\mu,4b$ channel, indicating that a softer background spectrum fits better for the latter. Therefore we do not assume preferred parameter values, and these parameters are determined each time by fitting to the $e^\pm$ data sets. 

Following~\citet{Barger:2009yt}, we first obtain the propagated Galactic and signal $e^\pm$ spectra on a grid on the five propagation and injection parameters. Then we interpolate the total $e^\pm$ and position fraction spectra to fit the combined $e^\pm$ data. The best-fit spectra are shown later in Fig.~\ref{fig:best_fit} and \ref{fig:best_fit_mixed}, and the goodness of fits are listed in Table~\ref{tab:fit}. 
 
First we consider the combination of {\it DAMPE}, Fermi-LAT and H.E.S.S. electron data. The TeV-range spectral shapes in these three data sets are in good agreement with each other. The number of data points is 71 and there are 7 free parameters in the fit: five from diffusion and background parametrization, one for H.E.S.S. energy scaling, plus one for the signal strength. As shown in Fig.~\ref{fig:best_fit} and Table~\ref{tab:fit}, the benchmark DM$\rightarrow 4\mu$ cascade decay gives a $2.3\sigma$ fit with 6 TeV mass, in case H.E.S.S. energy upscale by $\Delta f=$30\% to perfectly align with {\it DAMPE}, Fermi-LAT data points. The best-fit DM decay lifetime $2.0\times 10^{26}$s is allowed gamma ray searches. This scaling factor, however, is larger than the 15\% energy scale uncertainty in the H.E.S.S. experiment. Restricting $\Delta f\le 15\%$ leads to a much worse 4.3$\sigma$ fit, which indicated a minor normalization gap between the datasets. The non-scaled case $f=1$ is exclude by more than $8\sigma$.

The best $4\mu$ fit at $2.3\sigma$ consistency level indicate minor tension between the datasets in addition to the H.E.S.S. scaling factor. Note difference exists between {\it DAMPE} and Fermi-LAT data above 1 TeV, where the Fermi-LAT's electron spectrum does not fall as quickly as those in {\it DAMPE} and H.E.S.S. data. If only the {\it DAMPE} and H.E.S.S data are considered, which fully agree with each other's spectral features, a 1.2$\sigma$ fit can be achieved. Similar to the $4\mu$ channel, the three-step $8\mu$ and a mixed $4\mu,4b$ channels have softer $e^\pm$ injection spectra and also obtain good fits. The $4b$ component represents mixture into four-quark decay channels.

\begin{table*}
\begin{tabular}{c||c|c||c|c}
\hline
\hline
Data & 3 sets, $\Delta f<15\%$\ &\ 4 sets, $\Delta f<15\%$\ & 3 sets & 4 sets \\
\hline
DM(6 TeV)$\rightarrow 4\mu$ &1.9 (4.3$\sigma$)&  3.4 ($>$8$\sigma$)&  1.4 (2.3$\sigma$)& 2.9 ($>$8$\sigma$)\\
\hline
DM(10 TeV)$\rightarrow 8\mu$ &1.6 (3.2$\sigma$)&  2.5 (7.0$\sigma$)&  1.1 (1.1$\sigma$)& 2.0 (5.0$\sigma$)\\
\hline
DM(7 TeV)$\rightarrow 4\mu,4b$ &1.7 (3.4$\sigma$)&  1.7 (3.8$\sigma$)&  1.1 (1.3$\sigma$)& 1.2 (1.6$\sigma$)\\
\hline
\hline
\end{tabular}
\caption{\normalsize Reduced $\chi^2/$d.o.f. at joint three-dataset ({\it DAMPE}+FermiHE+H.E.S.S.) and four-dataset (with AMS02 positron fraction) best fits. The optimal DM mass is shown for each channel. The numbers inside parentheses show the goodness of fit, and we list both the results with restricted H.E.S.S. energy scaling range $\Delta f< 15\%$ and those with a free $f$ range.
The number of degrees of freedom is 65(80) with 3(4) data sets for the $4\mu$ and $8\mu$ channels. The mixed $4\mu,4b$ channel has one less degree of freedom due to varying the $4\mu$, $4b$ branching ratio.}
\label{tab:fit}
\end{table*}
\normalsize

We then add the AMS02 positron fraction data to the joint fit and the total number of data points increases to  86. While we do not include the AMS02 total electron spectrum in order to avoid major conflict between data sets, AMS02 still has the best precision measurement of the positron spectrum,  which is expected to originate from non-background contributions. In this case, the joint analysis does not produce a good fit with the muonic channels. The AMS02 data would prefer a much lower injection energy, namely the DM mass lower than that by the TeV spectrum in {\it DAMPE} and H.E.S.S. With the 6 TeV DM mass, the $4\mu$'s positron spectrum does not agree with the AMS02's positron data, and causes a significant rise in the total $\chi^2$. 

For comparison, the three-step $8\mu$ channel has a softer low-energy $e^\pm$ spectrum that helps to fit AMS02 data, yet its shape is not sufficiently softened to perfectly fit AMS02. A rather poor 5$\sigma$ fit is obtained for an 8 TeV DM with a lifetime $\tau_{\text{DM}}=$\sci{1.6}{26}s. It is clear from the spectral shape that the AMS02 data require a further softened positron spectrum in the ${\cal O}(10^{-3}-10^{-2})$ range of the maximal injection energy. 

A mixture into hadronic channel(s) can provide such a spectral feature. Fig.~\ref{fig:best_fit_mixed} shows a best-fit 7 TeV case of the mixed $4\mu,4b$ channel to the combined data sets and the overall goodness of fit over the DM mass range. Note for the mixed channel, the relative fraction between $4b,4\mu$ components is an additional free parameter. The best fit is obtained at $\chi^2=95$ out of 79 degrees of freedom, which indicates for $95\%$ confidence level consistency in a small mass range near 7-8 TeV if $f$ is allowed to upscale by 30\%, and the decay width is 0.5-0.8$\times 10^{26}$ s. When the energy scaling is restricted $\Delta f\le 15\%$ the optimal fit is still excluded by more than 3$\sigma$.

Here we should emphasize that while the hadronic component helps fitting the $e^{\pm}$ data, its high fraction (60-80\% at the best-fits) is gamma-ray excluded due to their $\pi^0$ production. The DM lifetime constraint for a pure 4$b$ cascade decay is to be longer than $10^{27}$s by isotropic diffuse gamma ray searches. This bound excludes the $4b$ branching fraction from exceeding a $1\%-2$\% level at the best-fit DM lifetime.  Within such a limited branching fraction, $4b$ can not qualitatively soften spectral shape to fit AMS02 data. Nevertheless, these results indicate AMS02's preference for a softer injection spectrum (than DM$\rightarrow 4\mu, 8\mu$) in the DM cascade decay scenario.

\section{Conclusion}
\label{sec:conclude}

With the assumption of a power-law like Galactic electron background, the cascade decay of multi-TeV DM into four leptons can accommodate the TeV-scale shape of cosmic ray electrons as measured by {\it DAMPE}, H.E.S.S. and Fermi-LAT. In the low mediator mass limit, the DM cascade decay scenario has reduced associated gamma ray radiation and a softened injection $e^\pm$ spectrum. We considered a 6 TeV DM$\rightarrow 4\mu$ channel as a benchmark scenario because this channel is the least constrained by diffuse gamma ray and CMB measurements. A best-fitting at $2.3\sigma$ confidence level is achieved for the joint {\it DAMPE}, H.E.S.S. and Fermi-LAT data, and a three-step $8\mu$ cascade improves the fit to 1.1$\sigma$. Such fits require H.E.S.S. data to up-float beyond its typical 15\% uncertainty range. Restricted energy-scaling fits with $4\mu, 8\mu$ cascades do not yield fits within $3\sigma$. After including the AMS02 positron fraction data, the muonic channels are disfavored. A gamma-ray constrained $4\mu,4b$ mixed channel is then considered and it is possible to fit all four data sets at the cost of 30\% upscaling in H.E.S.S. energy scale. This demonstrates that in the fit to the latest TeV cosmic electron data sets, AMS02 strongly prefers a very softened injection spectrum in the assumption of a power-law like Galactic background. Further study for viable cascade decay processes will help us to understand the decaying dark matter explanation for the TeV structure in the cosmic ray electron spectrum from {\it DAMPE} and H.E.S.S. observations. 

\vskip 0.1 truein

\noindent \textbf{Acknowledgments:} Y.G. is supported under grant no.~Y7515560U1 by the Institute of High Energy Physics, Chinese Academy of Sciences. Y.Z.M is supported by the National Research Foundation of South Africa with Grant no.~105925 and no.~104800, and National Science Foundation of China with Grant no.~11828301.

\appendix

\bibliographystyle{mn2e}

\bibliography{refs}

\begin{thebibliography}{54}
\expandafter\ifx\csname natexlab\endcsname\relax\def\natexlab#1{#1}\fi

\bibitem[{{Abdo} \& {et.al.}(2009)}]{bib:fermi_e}
{Abdo} A.~A., {et.al.}, 2009, Physical Review Letters, 102, 181101

\bibitem[{Abdollahi {et~al.}(2017)}]{Abdollahi:2017nat}
Abdollahi S., {et~al.}, 2017, Phys. Rev., D95, 082007

\bibitem[{{Ackermann} {et~al.}(2015){Ackermann}, {et.al. }, \& {Fermi-LAT
  Collaboration}}]{bib:fermi_gamma}
{Ackermann} M., {et.al. }, {Fermi-LAT Collaboration}, 2015, Physical Review
  Letters, 115, 231301

\bibitem[{Ackermann {et~al.}(2017)}]{TheFermiLAT:2017vmf}
Ackermann M., {et~al.}, 2017, Astrophys. J., 840, 43

\bibitem[{{Adapted from Samuel Ting's CERN Colloquium, on 24
  May,}(2018)}]{TingCernAMStalk}
{Adapted from Samuel Ting's CERN Colloquium, on 24 May,}, 2018

\bibitem[{{Adriani} \& {et.al.}(2011)}]{bib:pamela}
{Adriani} O., {et.al.}, 2011, Physical Review Letters, 106, 201101

\bibitem[{{Adriani} {et~al.}(2010){Adriani}, {et.al.}, \& {PAMELA
  Collaboration}}]{bib:pamela_antiproton}
{Adriani} O., {et.al.}, {PAMELA Collaboration}, 2010, Physical Review Letters,
  105, 121101

\bibitem[{{Aguilar} {et~al.}(2013){Aguilar}, {Alberti}, {Alpat}, {Alvino},
  {Ambrosi}, {Andeen}, {Anderhub}, {Arruda}, {Azzarello}, {Bachlechner}, \&
  et~al.}]{bib:ams02}
{Aguilar} M., {Alberti} G., {Alpat} B., {Alvino} A., {Ambrosi} G., {Andeen} K.,
  {Anderhub} H., {Arruda} L., {Azzarello} P., {Bachlechner} A., et~al., 2013,
  Physical Review Letters, 110, 141102

\bibitem[{{Aguilar} {et~al.}(2016){Aguilar}, {et.al.}, \& {AMS
  Collaboration}}]{bib:ams02_antiproton}
{Aguilar} M., {et.al.}, {AMS Collaboration}, 2016, Phys. Rev. Lett., 117,
  231102

\bibitem[{Aguilar {et~al.}(2015)}]{Aguilar:2015ctt}
Aguilar M., {et~al.}, 2015, Phys. Rev. Lett., 115, 211101

\bibitem[{Aguilar {et~al.}(2016)}]{Aguilar:2016kjl}
---, 2016, Phys. Rev. Lett., 117, 091103

\bibitem[{{Aharonian} \& {et.al.}(2009)}]{bib:hess}
{Aharonian} F., {et.al.}, 2009, \aap, 508, 561

\bibitem[{{Allahverdi} {et~al.}(2009){Allahverdi}, {Dutta},
  {Richardson-McDaniel}, \& {Santoso}}]{Allahverdi:2009ae}
{Allahverdi} R., {Dutta} B., {Richardson-McDaniel} K., {Santoso} Y., 2009,
  Physics Letters B, 677, 172

\bibitem[{{Anisimov} \& {di Bari}(2009)}]{2009PhRvD..80g3017A}
{Anisimov} A., {di Bari} P., 2009, \prd, 80, 073017

\bibitem[{{Arvanitaki} {et~al.}(2009){Arvanitaki}, {Dimopoulos}, {Dubovsky},
  {Graham}, {Harnik}, \& {Rajendran}}]{2009PhRvD..79j5022A}
{Arvanitaki} A., {Dimopoulos} S., {Dubovsky} S., {Graham} P.~W., {Harnik} R.,
  {Rajendran} S., 2009, \prd, 79, 105022

\bibitem[{{Bahcall} \& {Soneira}(1980)}]{IsoT}
{Bahcall} J.~N., {Soneira} R.~M., 1980, \apjs, 44, 73

\bibitem[{{Barger} {et~al.}(2009){Barger}, {Gao}, {Keung}, {Marfatia}, \&
  {Shaughnessy}}]{Barger:2009yt}
{Barger} V., {Gao} Y., {Keung} W.-Y., {Marfatia} D., {Shaughnessy} G., 2009,
  Physics Letters B, 678, 283

\bibitem[{{Biermann} {et~al.}(2009){Biermann}, {Becker}, {Meli}, {Rhode},
  {Seo}, \& {Stanev}}]{bib:Biermann09}
{Biermann} P.~L., {Becker} J.~K., {Meli} A., {Rhode} W., {Seo} E.~S., {Stanev}
  T., 2009, Physical Review Letters, 103, 061101

\bibitem[{{Blasi}(2009)}]{bib:Blasi09}
{Blasi} P., 2009, Physical Review Letters, 103, 051104

\bibitem[{{Bullock} \& {Boylan-Kolchin}(2017)}]{bib:Bullock17}
{Bullock} J.~S., {Boylan-Kolchin} M., 2017, \araa, 55, 343

\bibitem[{Carlson {et~al.}(2016)Carlson, Linden, \& Profumo}]{Carlson:2016iis}
Carlson E., Linden T., Profumo S., 2016, Phys. Rev., D94, 063504

\bibitem[{{Chang} \& {et.al.}(2008)}]{bib:atic}
{Chang} J., {et.al.}, 2008, \nat, 456, 362

\bibitem[{{Chao} {et~al.}(2018){Chao}, {Guo}, {Li}, \& {Shu}}]{Chao:2017emq}
{Chao} W., {Guo} H.-K., {Li} H.-L., {Shu} J., 2018, Physics Letters B, 782, 517

\bibitem[{Cholis {et~al.}(2009)Cholis, Goodenough, \& Weiner}]{Cholis:2008vb}
Cholis I., Goodenough L., Weiner N., 2009, Phys. Rev., D79, 123505

\bibitem[{{Clark} {et~al.}(2018){Clark}, {Dutta}, {Gao}, {Ma}, \&
  {Strigari}}]{2018PhRvD..98d3006C}
{Clark} S.~J., {Dutta} B., {Gao} Y., {Ma} Y.-Z., {Strigari} L.~E., 2018, \prd,
  98, 043006

\bibitem[{Clark {et~al.}(2018)Clark, Dutta, \& Strigari}]{Clark:2017fum}
Clark S.~J., Dutta B., Strigari L.~E., 2018, Phys. Rev., D97, 023003

\bibitem[{{Cui} {et~al.}(2017){Cui}, {Yuan}, {Tsai}, \& {Fan}}]{Cui17}
{Cui} M.-Y., {Yuan} Q., {Tsai} Y.-L.~S., {Fan} Y.-Z., 2017, Physical Review
  Letters, 118, 191101

\bibitem[{{DAMPE Collaboration} {et~al.}(2017){DAMPE Collaboration}, {Ambrosi},
  \& {et.al.}}]{DAMPE}
{DAMPE Collaboration}, {Ambrosi} G., {et.al.}, 2017, \nat, 552, 63

\bibitem[{{Duan} {et~al.}(2018){Duan}, {Feng}, {Wang}, {Wu}, {Yang}, \&
  {Zheng}}]{Duan17}
{Duan} G.~H., {Feng} L., {Wang} F., {Wu} L., {Yang} J.~M., {Zheng} R., 2018,
  Journal of High Energy Physics, 2, 107

\bibitem[{Duerr {et~al.}(2016)Duerr, Fileviez~Pérez, \&
  Smirnov}]{Duerr:2015bea}
Duerr M., Fileviez~Pérez P., Smirnov J., 2016, JHEP, 06, 008

\bibitem[{{Einasto}(1965)}]{Einasto65}
{Einasto} J., 1965, Trudy Astrofizicheskogo Instituta Alma-Ata, 5, 87

\bibitem[{{Fan} {et~al.}(2018){Fan}, {Huang}, {Spinrath}, {Tsai}, \&
  {Yuan}}]{Fan17}
{Fan} Y.-Z., {Huang} W.-C., {Spinrath} M., {Tsai} Y.-L.~S., {Yuan} Q., 2018,
  Physics Letters B, 781, 83

\bibitem[{{Fang} {et~al.}(2018){Fang}, {Bi}, \& {Yin}}]{Fang:2017tvj}
{Fang} K., {Bi} X.-J., {Yin} P.-F., 2018, \apj, 854, 57

\bibitem[{{Feng}(2010)}]{Feng10}
{Feng} J.~L., 2010, \araa, 48, 495

\bibitem[{{Goodenough} \& {Hooper}(2009)}]{Goodenough:2009gk}
{Goodenough} L., {Hooper} D., 2009, arXiv e-prints, arXiv:0910.2998

\bibitem[{{Hooper} {et~al.}(2009){Hooper}, {Blasi}, \& {Dario
  Serpico}}]{bib:pulsar}
{Hooper} D., {Blasi} P., {Dario Serpico} P., 2009, \jcap, 1, 025

\bibitem[{{Kerszberg} {et~al.}(2017){Kerszberg}, {Kraus}, {Kolitzus},
  {Egberts}, {Funk}, {Lenain}, {Reimer}, \& {Vincent}}]{HESS2017}
{Kerszberg} D., {Kraus} M., {Kolitzus} D., {Egberts} K., {Funk} S., {Lenain}
  J.-P., {Reimer} O., {Vincent} P., 2017, Contributions of the High Energy
  Stereoscopic System (H.E.S.S.) to the 35th International Cosmic Ray
  Conference (ICRC), Busan, Korea

\bibitem[{{Liu} {et~al.}(2017{\natexlab{a}}){Liu}, {Bi}, {Lin}, {Wang}, \&
  {Yin}}]{bib:remnants}
{Liu} W., {Bi} X.-J., {Lin} S.-J., {Wang} B.-B., {Yin} P.-F.,
  2017{\natexlab{a}}, \prd, 96, 023006

\bibitem[{{Liu} {et~al.}(2017{\natexlab{b}}){Liu}, {Bi}, {Lin}, \&
  {Yin}}]{Liu:2016ngs}
{Liu} W., {Bi} X.-J., {Lin} S.-J., {Yin} P.-F., 2017{\natexlab{b}}, Chinese
  Physics C, 41, 045104

\bibitem[{Malyshev(2017)}]{Malyshev:2017pmr}
Malyshev D., 2017, PoS, IFS2017, 052

\bibitem[{{Malyshev} {et~al.}(2009){Malyshev}, {Cholis}, \&
  {Gelfand}}]{bib:Malyshev09}
{Malyshev} D., {Cholis} I., {Gelfand} J., 2009, \prd, 80, 063005

\bibitem[{Moskalenko {et~al.}(2006)Moskalenko, Porter, \&
  Strong}]{Moskalenko:2005ng}
Moskalenko I.~V., Porter T.~A., Strong A.~W., 2006, Astrophys. J., 640, L155

\bibitem[{Panov {et~al.}(2009)}]{Panov:2011ak}
Panov A.~D., {et~al.}, 2009, Bull. Russ. Acad. Sci. Phys., 73, 564

\bibitem[{{Planck Collaboration} {et~al.}(2016){Planck Collaboration}, {Ade},
  {Aghanim}, {Arnaud}, {Ashdown}, {Aumont}, {Baccigalupi}, {Banday},
  {Barreiro}, {Bartlett}, \& et~al.}]{bib:planck}
{Planck Collaboration}, {Ade} P.~A.~R., {Aghanim} N., {Arnaud} M., {Ashdown}
  M., {Aumont} J., {Baccigalupi} C., {Banday} A.~J., {Barreiro} R.~B.,
  {Bartlett} J.~G., et~al., 2016, \aap, 594, A13

\bibitem[{{Profumo} {et~al.}(2018){Profumo}, {Queiroz}, {Silk}, \&
  {Siqueira}}]{Profumo:2017obk}
{Profumo} S., {Queiroz} F.~S., {Silk} J., {Siqueira} C., 2018, \jcap, 2018, 010

\bibitem[{{Shaviv} {et~al.}(2009){Shaviv}, {Nakar}, \& {Piran}}]{bib:Shaviv09}
{Shaviv} N.~J., {Nakar} E., {Piran} T., 2009, Physical Review Letters, 103,
  111302

\bibitem[{{Slatyer} \& {Wu}(2017)}]{Slatyer:2016qyl}
{Slatyer} T.~R., {Wu} C.-L., 2017, \prd, 95, 023010

\bibitem[{{Strong} \& {Moskalenko}(1999)}]{Strong:1999sv}
{Strong} A.~W., {Moskalenko} I.~V., 1999, International Cosmic Ray Conference,
  4, 255

\bibitem[{{Strong} \& {Moskalenko}(2001)}]{bib:Strong01}
---, 2001, International Cosmic Ray Conference, 5, 1942

\bibitem[{{Strong} {et~al.}(2007){Strong}, {Moskalenko}, \&
  {Ptuskin}}]{Strong:2007nh}
{Strong} A.~W., {Moskalenko} I.~V., {Ptuskin} V.~S., 2007, Annual Review of
  Nuclear and Particle Science, 57, 285

\bibitem[{Yoon {et~al.}(2017)}]{Yoon:2017qjx}
Yoon Y.~S., {et~al.}, 2017, Astrophys. J., 839, 5

\bibitem[{{Yuan} {et~al.}(2017){Yuan}, {Feng}, {Yin}, {Fan}, {Bi}, {Cui},
  {Dong}, {Guo}, {Fang}, {Hu}, {Huang}, {Lei}, {Li}, {Lin}, {Liu}, {Ma},
  {Peng}, {Qiao}, {Shen}, {Su}, {Wei}, {Xu}, {Yue}, {Zang}, {Zhang}, {Zhang},
  {Zhang}, {Zhang}, \& {Zhang}}]{Yuan17}
{Yuan} Q., {Feng} L., {Yin} P.-F., {Fan} Y.-Z., {Bi} X.-J., {Cui} M.-Y., {Dong}
  T.-K., {Guo} Y.-Q., {Fang} K., {Hu} H.-B., {Huang} X., {Lei} S.-J., {Li} X.,
  {Lin} S.-J., {Liu} H., {Ma} P.-X., {Peng} W.-X., {Qiao} R., {Shen} Z.-Q.,
  {Su} M., {Wei} Y.-F., {Xu} Z.-L., {Yue} C., {Zang} J.-J., {Zhang} C., {Zhang}
  X., {Zhang} Y.-P., {Zhang} Y.-J., {Zhang} Y.-L., 2017, arXiv e-prints,
  arXiv:1711.10989

\bibitem[{{Y{\"u}ksel} {et~al.}(2009){Y{\"u}ksel}, {Kistler}, \&
  {Stanev}}]{bib:Yuksel09}
{Y{\"u}ksel} H., {Kistler} M.~D., {Stanev} T., 2009, Physical Review Letters,
  103, 051101

\bibitem[{Zhou {et~al.}(2015)Zhou, Liang, Huang, Li, Fan, Feng, \&
  Chang}]{Zhou:2014lva}
Zhou B., Liang Y.-F., Huang X., Li X., Fan Y.-Z., Feng L., Chang J., 2015,
  Phys. Rev., D91, 123010

\end{thebibliography}

\end{document}